\documentclass[12pt]{article}
\usepackage{epsf}

\begin{document}

\begin{titlepage}
\begin{flushright}
CLNS~03/1814\\
{\tt hep-ph/0212360}\\[0.2cm]
December 24, 2002
\end{flushright}

\vspace{1.0cm}
\begin{center}
\Large\bf Radiative B Decays:\\
Standard Candles of Flavor Physics\footnote{Invited plenary talk presented 
at the 10th International Conference on {\em Supersymmetry and 
Unification}, DESY, Hamburg (17--23 June 2002)}
\end{center}

\vspace{1.0cm}
\begin{center}
Matthias Neubert\\[0.1cm]
{\sl F.R. Newman Laboratory for Elementary-Particle Physics\\
Cornell University, Ithaca, NY 14853, USA}
\end{center}

\vspace{1.0cm}
\begin{abstract}
\vspace{0.2cm}\noindent 
Rare radiative decays based on $b\to s\gamma$ transitions are among the 
most prominent examples of flavor-changing neutral current processes. 
They benefit from good theoretical control and experimental 
accessibility, large sensitivity to physics beyond the Standard Model, 
and the availability of many observables. In this talk I summarize the 
status of the theoretical understanding of these decays and review how 
they may be used to constrain extensions of the Standard Model, with 
particular focus on supersymmetric models.
\end{abstract}
\vfill
\end{titlepage}

\section{Introduction}

\begin{figure}[ht]
\epsfxsize=7cm
\centerline{\epsffile{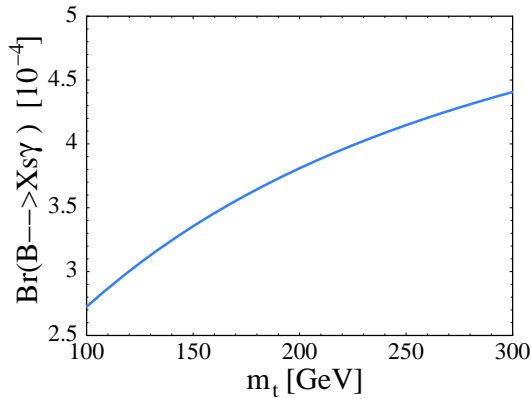}}
\centerline{\parbox{15cm}{\caption{\label{fig:mtdep}
Sensitivity of the $B\to X_s\gamma$ branching ratio to the top-quark 
mass.}}}
\end{figure}

Rare radiative decays of $B$ mesons mediated by the quark decay 
$b\to s\gamma$ are prime examples of flavor-changing neutral current 
(FCNC) transitions. They are forbidden in the Standard Model (SM) at tree 
level and so are sensitive to the contributions of heavy particles in 
loop diagrams. Figure~\ref{fig:mtdep} illustrates this fact for the case 
of the SM, showing the strong dependence of the inclusive $B\to X_s\gamma$
branching ratio on the mass of the top quark.

$B\to X_s\gamma$ decays are an excellent probe for physics beyond the SM
because their rate is small yet well measured experimentally, this rate 
can be calculated with high precision, and it shows large sensitivity to 
non-standard sources of flavor violation and CP violation. Thus, the 
study of these decays provides powerful constraints on many New Physics 
scenarios, including models with supersymmetry (SUSY).

\section{Inclusive \boldmath$B\to X_s\gamma$ Decay Rate\unboldmath}

\begin{figure}
\epsfxsize=12cm
\centerline{\epsffile{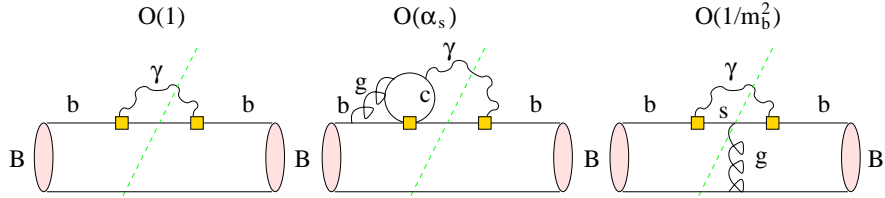}}
\centerline{\parbox{15cm}{\caption{\label{fig:OPE}
Application of the operator product expansion to the calculation of
inclusive $B$-meson decay rates.}}}
\end{figure}

Starting point of the most sophisticated calculation in flavor physics is
the effective weak Hamiltonian
\[
   H_{\rm eff} = - \frac{G_F}{\sqrt2}\,V_{tb} V_{ts}^*\,
   \sum_i C_i(\mu)\,Q_i(\mu) \,.
\]
Information about New Physics and heavy particles is encoded in the
short-distance (Wilson) coefficient functions $C_i$, while the hadronic 
matrix elements of the operators $Q_i$ contain all long-distance 
strong-interaction effects. The evaluation of these matrix elements 
constitutes the principal theoretical challenge in obtaining precise 
predictions for the decay rate. A consistent calculation of the Wilson 
coefficients at next-to-leading order (NLO) requires 3-loop anomalous 
dimensions \cite{Chetyrkin:1996vx}, electroweak radiative corrections 
\cite{Czarnecki:1998tn,Kagan:1998ym,Gambino:2001au}, and 2-loop matching 
conditions at the weak scale. These matching conditions depend on the 
underlying high-energy theory and so are sensitive to physics beyond the 
SM. They are known for the SM \cite{Adel:1993ah,Greub:1997hf}, 
two-Higgs-doublet models (2HDMs) \cite{Ciuchini:1997xe,Borzumati:1998tg}, 
left-right symmetric models \cite{Bobeth:1999ww}, the so-called 
``constrained minimal supersymmetric SM'' (CMSSM) 
\cite{Bobeth:1999ww,Ciuchini:1998xy}, and the CMSSM with large $\tan\beta$
\cite{Degrassi:2000qf,Carena:2000uj}. The relevant hadronic matrix 
elements required for the total inclusive $B\to X_s\gamma$ decay rate are 
calculated using the operator product expansion, as illustrated in 
Figure~\ref{fig:OPE}. At NLO one needs 2-loop matrix elements of 
four-quark operators \cite{Greub:1996jd,Buras:2001mq}. It is state of the 
art to include power corrections of order $(\Lambda_{\rm QCD}/m_b)^2$ and 
$(\Lambda_{\rm QCD}/m_c)^2$ in the heavy-quark expansion, using the 
techniques of heavy-quark effective theory 
\cite{Neubert:1993mb,Falk:1993dh,Voloshin:1996gw}.

\begin{figure}[ht]
\epsfxsize=8cm
\centerline{\epsffile{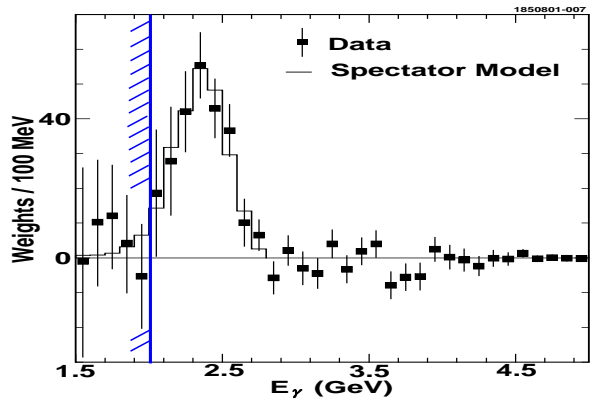}}
\centerline{\parbox{15cm}{\caption{\label{fig:Fermi}
Cut on the photon energy applied in the CLEO analysis of the 
$B\to X_s\gamma$ branching ratio \cite{Chen:2001fj}.}}}
\end{figure}

Measurements of the inclusive $B\to X_s\gamma$ decay rate rely on a cut 
on the photon energy in the $B$-meson rest frame, because only highly 
energetic photons can be distinguished from the background. Typically, 
only events with $E_\gamma>2.0$\,GeV or so are recorded, as shown in 
Figure~\ref{fig:Fermi}. Accounting for such a cut theoretically is 
difficult and introduces sensitivity to the shape of the photon spectrum 
(``Fermi motion''), which can be analyzed using the twist expansion 
\cite{Neubert:1993ch,Bigi:1993ex,Neubert:1993um}.

Recently, there have been several improvements in the theoretical 
understanding of the inclusive $B\to X_s\gamma$ branching ratio. It has 
been pointed out that, because the photon-energy cut ensures that no open 
charm can be produced in the final state, it is appropriate to use a 
running charm-quark mass rather than the pole mass in the calculation of 
diagrams containing charm-quark loops \cite{Gambino:2001ew}. This leads 
to an enhancement of the rate by about 10\%. The calculation of the 
2-loop matrix elements for penguin operators has been completely, which 
however has a negligible effect on the rate \cite{Buras:2002tp}. Finally, 
it has been pointed out that one can avoid the common practice of 
normalizing the radiative rate to the semileptonic $B\to X_c l\nu$ rate 
(which introduces additional uncertainties) by using a physical $b$-quark 
mass definition instead of the pole mass \cite{Becher:2002ue}. The final 
results obtained with an (unrealistically low) energy cut 
$E_\gamma>1.6$\,GeV are
\[
   \mbox{Br}(B\to X_s\gamma) = \cases{
    (3.57\pm 0.30)\cdot 10^{-4} & \cite{Buras:2002tp} , \cr
    (3.54\pm 0.30)\cdot 10^{-4} & \cite{Becher:2002ue} . \cr}
\]
Extrapolation of these values to $E_\gamma>2.0$\,GeV yields 
\cite{Becher:2002ue}
\[
   \mbox{Br}(B\to X_s\gamma)
   = (3.26\pm 0.27_{\,-0.18}^{\,+0.09})\cdot 10^{-4} \,.
\]
Here the second error accounts for the uncertainty in the treatment of 
Fermi motion \cite{Kagan:1998ym}. The theoretical value compares well 
with the CLEO measurement 
$\mbox{Br}(B\to X_s\gamma)=(2.94\pm 0.39\pm 0.25)\cdot 10^{-4}$ 
\cite{Chen:2001fj} obtained with $E_\gamma>2.0$\,GeV. One should refrain 
from extrapolating the theoretical predictions down to much lower 
energies, where in any case no measurement can be done. This 
extrapolation is plagued by large theoretical uncertainties in the 
treatment of soft photons and $c\bar c$ resonance production. 
Unfortunately, the Belle collaboration has chosen to extrapolate their 
measurement to lower energies using a theoretical model. Their result
$(3.36\pm 0.53\pm 0.42_{\,-0.54}^{\,+0.50})\cdot 10^{-4}$ 
\cite{Abe:2001hk} may be compared with the theoretical extrapolation
$\mbox{Br}(B\to X_s\gamma) = (3.64\pm 0.31)\cdot 10^{-4}$ obtained from
the above results using the same model.

The most important conclusion to be drawn from the excellent agreement 
between SM theory and experimental data is that there appears to be no 
room left for drastic New-Physics effects in FCNC processes based on 
$b\to s\gamma$ transitions.

\section{Generic Implications for New Physics}

A large portion of the New-Physics literature focuses on models with 
``minimal flavor violation'', whose primary motivation is to avoid 
disasters in the flavor sector (see \cite{D'Ambrosio:2002ex} for an
elegant effective field-theory approach to this class of models). In such 
models the CKM matrix is assumed to be the only source of quark-flavor 
mixing. Prominent examples are the type-II 2HDM and the CMSSM. In these 
scenarios only moderate FCNC effects are allowed after the constraints 
from electroweak precision data are taken into account. Models with 
minimal flavor violation are phenomenologically ``preferred'', since data 
show no evidence for non-standard flavor or CP violation. However, these 
models are theoretically somewhat ad hoc.

More generic models of New Physics contain new sources of flavor 
violation. Examples are general SUSY extensions of the SM, models with 
new quark generations, etc. These models are more ``natural'', since 
after all we expect some physics beyond the SM to explain the origin of 
flavor. Generically, however, they can have drastic effects on FCNC 
processes, such as $K$--$\bar K$ mixing, $B\to X_s\gamma$, 
$K\to\pi\nu\bar\nu$, etc. In particular, generic SUSY models can naturally
lead to huge FCNC effects. (Or those effects were predicted before their 
existence was excluded experimentally.) The fact that such large effects 
are not realized in Nature is sometimes called the SUSY flavor problem. 

At this point a caveat is in order. Most extensions of the SM come with a 
plethora of new parameters, most of which are related to the flavor 
sector (e.g., there are 43 new CP-violating phases in the MSSM!). In the 
context of particle searches it is a common (perhaps legitimate) practice 
to make vastly simplifying assumptions about these parameters, typically 
reducing their number from over a hundred (in ``minimal'' SUSY models) to 
less than about 5. These simplifications are dangerous in the context of 
flavor physics. In a generic model, basically each flavor-changing 
process receives its own New-Physics contributions. Adjusting flavor 
parameters in an ad hoc way may lead to correlations between observables 
that are strongly model dependent (such as, e.g., correlations between 
$K$--$\bar K$ mixing 
$\leftrightarrow~B\to X_s\gamma~\leftrightarrow~K\to\pi\nu\bar\nu$).

\section{Specific New-Physics Models}

After these remarks, let me now discuss some particular New-Physics 
scenarios in more detail.

\subsection{Type-II 2HDM}

\begin{figure}
\epsfxsize=8cm
\centerline{\epsffile{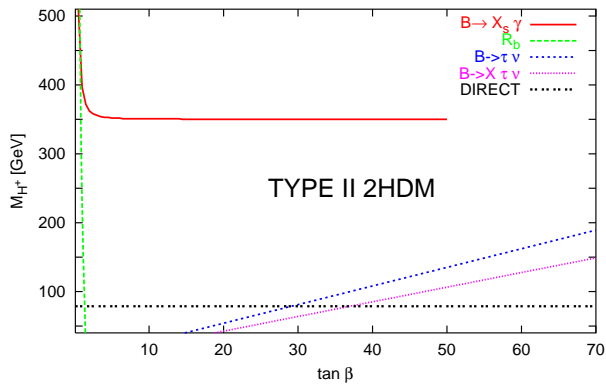}}
\centerline{\parbox{15cm}{\caption{\label{fig:higgs}
Bound on the charged-Higgs mass in the type-II 2HDM derived from the 
analysis of the inclusive $B\to X_s\gamma$ branching ratio (solid line)
\cite{Gambino:2001ew}.}}}
\end{figure}

In this model there is a charged-Higgs contribution to the $b\to s\gamma$ 
transition amplitude, which adds constructively to the SM contribution. 
As a result, one obtains a strong bound on the charged-Higgs mass, which 
is in fact stronger than the bounds obtained from direct searches. The 
complete NLO analysis of this bound has been presented in 
\cite{Ciuchini:1997xe,Borzumati:1998tg}. The most recent evaluation 
yields $m_{H^+}>350$\,GeV at 99\% CL \cite{Gambino:2001ew}, as 
illustrated in Figure~\ref{fig:higgs}.

More generally, one should expect constructive or destructive 
interference of New-Physics effects with the SM contribution. A useful 
general formula is \cite{Kagan:1998ym}
\begin{eqnarray}
   10^4\,\mbox{Br}(B\to X_s\gamma)
   &\approx& 3.26 + 1.40\,\mbox{Re}\,\xi_7 + 0.14\,\mbox{Re}\,\xi_8
    \nonumber\\
   &&\mbox{}+ 0.37 \left( |\xi_7|^2 + |\xi_7^R|^2 \right)
   + 0.08\,\mbox{Re} \left( \xi_7\xi_8^* + \xi_7^R\xi_8^{R*} \right) ,
   \nonumber
\end{eqnarray}
where the New-Physics contributions are parameterized as
\[
   \xi_{7,8} = \frac{C_{7,8}^{\rm NP}(m_W)}{C_{7,8}^{\rm SM}(m_W)} \,,
   \qquad
   \xi_{7,8}^R = \frac{C_{7,8}^{R,{\rm NP}}(m_W)}{C_{7,8}^{\rm SM}(m_W)}
   \,.
\]
Here $C_7$ and $C_8$ are the Wilson coefficients of the electro-magnetic
and chromo-magnetic dipole operators, respectively, and $C_{7,8}^R$ are
the corresponding coefficients of non-standard operators with the 
opposite chirality of the quark fields. Note that despite the apparent 
agreement between data and SM theory (corresponding to $\xi_i=0$) it is 
still possible to have significant New-Physics contributions in 
$B\to X_s\gamma$ decays, provided that Re$\,\xi_{7,8}<0$ (destructive 
interference) and one is willing to accept some moderate fine-tuning.

\subsection{CMSSM with minimal flavor violation}

\begin{figure}
\epsfxsize=12cm
\centerline{\epsffile{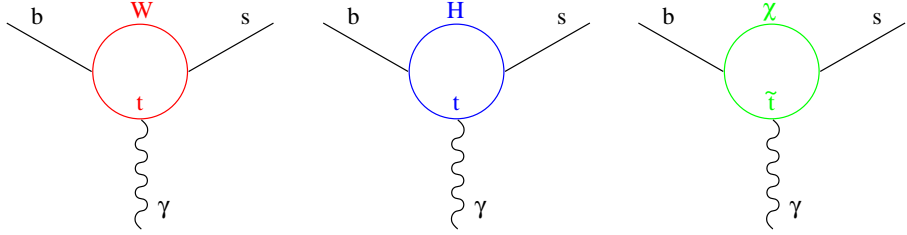}}
\centerline{\parbox{15cm}{\caption{\label{fig:CMSSM}
Examples of SUSY penguin diagrams present in the CMSSM.}}}
\end{figure}

\begin{figure}[ht]
\epsfxsize=8.5cm
\centerline{\epsffile{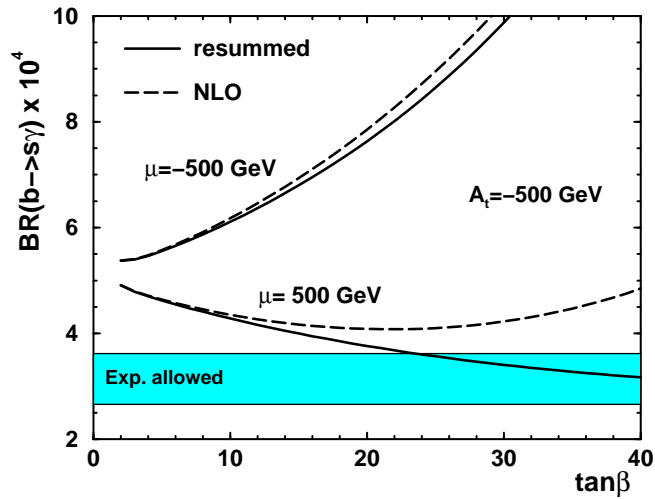}}
\centerline{\parbox{15cm}{\caption{\label{fig:marcela}
$B\to X_s\gamma$ constraints on the parameters of the CMSSM 
\cite{Carena:2000uj}.}}}
\end{figure}

In this highly constrained SUSY model there are three types of 
contributions to the dipole coefficients:
\[
   C_{7,8}(m_W) = \underbrace{C_{7,8}^{\rm SM}(m_W)}_{\mbox{\tiny SM}}
   + \underbrace{C_{7,8}^{\rm H}(m_W)}_{\mbox{\tiny type-II 2HDM}}
   + \underbrace{C_{7,8}^{\chi}(m_W)}_{\mbox{\tiny chargino-stop}}
\]
They are illustrated in Figure~\ref{fig:CMSSM}. Several recent analyses 
of these contributions exist, some including novel higher-order terms 
that are enhanced for large $\tan\beta$ 
\cite{Degrassi:2000qf,Carena:2000uj,Demir:2001yz,Boz:2002wa}. As shown in 
Figure~\ref{fig:marcela}, agreement with the data strongly favors 
negative values of $\mu A_t$ (with positive $\mu$). An important finding 
is that large-$\tan\beta$ corrections can weaken the bound on the 
charged-Higgs mass significantly, even in the decoupling limit.

\subsection{Unconstrained MSSM}

\begin{figure}
\epsfxsize=15cm
\centerline{\epsffile{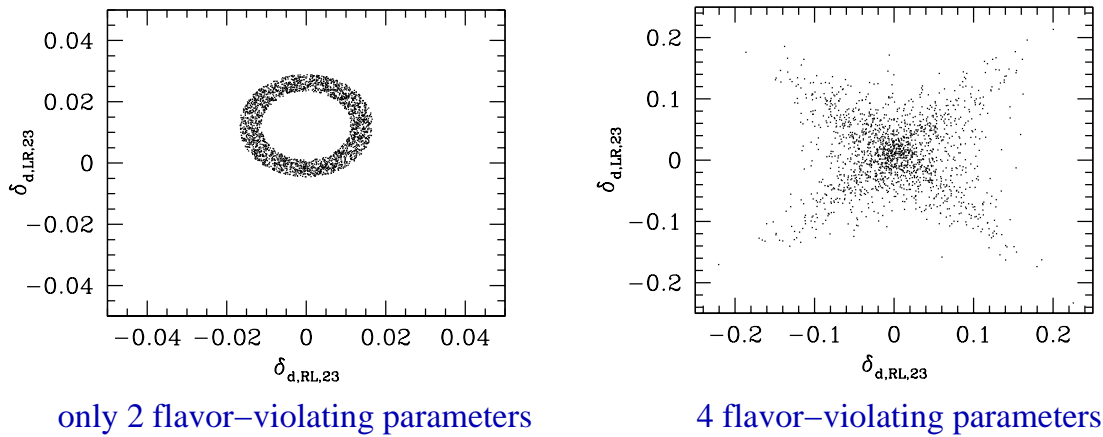}}
\centerline{\parbox{15cm}{\caption{\label{fig:Greub}
Constraints on SUSY flavor violation in a general scenario without minimal
flavor violation \cite{Besmer:2001cj}.}}}
\end{figure}

In a more general (but still ``minimal'') SUSY scenario there are new 
flavor-changing quark-squark-gluino couplings, which can be parameterized
in terms of off-diagonal entries in the squark mass matrix, e.g.
$\delta_{23}^{LR}=(m_{LR}^2)_{23}/m_{\tilde q}^2$, which by naive power
counting are expected to be of $O(1)$. Many analyses of such couplings 
have adopted the mass-insertion approximation (see, e.g., 
\cite{Hagelin:1992tc,Gabbiani:1996hi}), accompanied by the simplifying 
assumption that a single flavor-changing coupling is responsible for the 
dominant New-Physics effects. Recently, a more complete analysis of SUSY 
flavor violation taking into account the interplay of contributions from 
gluinos, neutralinos, charginos, and charged Higgs has been presented 
\cite{Besmer:2001cj}. As illustrated in Figure~\ref{fig:Greub}, the 
authors find that the resulting constraints on $\delta_{23}^{LR,RL}$ can 
be significantly relaxed (typically by an order of magnitude) due to 
interference effects.

\subsection{Flavor violation from light \boldmath$\tilde b$ 
squarks\unboldmath}

\begin{figure}
\epsfxsize=11cm
\centerline{\epsffile{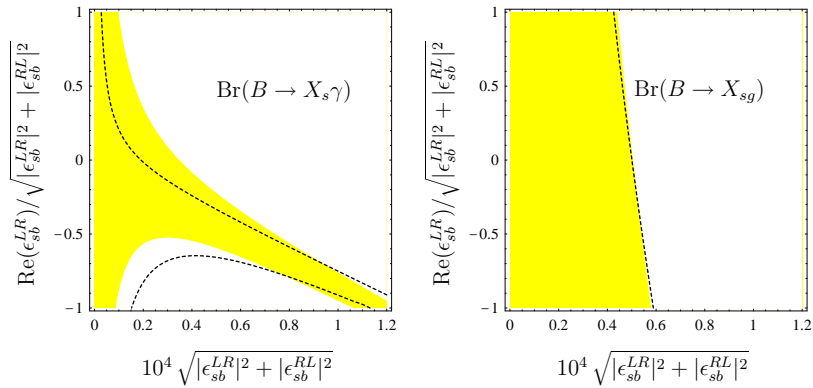}}
\centerline{\parbox{15cm}{\caption{\label{fig:bsq}
Constraints on the flavor-changing couplings describing SUSY flavor 
violations mediated by a light $\tilde b$ squark \cite{Becher:2002ue}.}}}
\end{figure}

The presence of a light $\tilde b$ squark with mass $\sim 2$--4 GeV,
accompanied by a light gluino with mass $\sim 15$\,GeV, could explain the 
observed excess of $b$-production at the Tevatron \cite{Berger:2000mp}.
This would naturally give rise to new sources of $b\to s$ FCNC 
transitions. The resulting flavor violations can be parameterized in 
terms of parameters $\epsilon_{sb}^{LR}$ etc., which naively could be of 
$O(1)$. However, a complete NLO analysis of the inclusive 
$B\to X_s\gamma$ branching ratio in this scenario yields extremely tight 
constraints on these couplings \cite{Becher:2002ue}, as illustrated in 
Figure~\ref{fig:bsq}. This imposes severe constraints on model building.

\section{CP Asymmetry in \boldmath$B\to X_s\gamma$ Decays\unboldmath}

Searching for direct CP violation in radiative decays provides an 
additional, powerful probe for physics beyond the SM
\cite{Soares:1991te,Wolfenstein:1994jw,Asatrian:1996as,Kagan:1998bh}. This
is basically a null effect in the SM, since
\[
   A_{\rm CP}^{\rm SM}(B\to X_s\gamma)\sim 
   \underbrace{\alpha_s(m_b)}_{\mbox{\tiny strong phase}}
   \!\!\times\!\!
   \underbrace{\frac{V_{ub}}{V_{cb}}}_{\mbox{\tiny CKM suppr.}}
   \!\!\times\!\!
   \underbrace{\frac{m_c^2}{m_b^2}}_{\mbox{\tiny GIM suppr.}}
   \approx 0.5\% \,.
\]
Moreover, in the SM the asymmetry vanishes (due to unitarity of the CKM 
matrix) if  no distinction between $s$ and $d$ quarks in the final state 
is made. 

\begin{figure}[ht]
\epsfxsize=7cm
\centerline{\epsffile{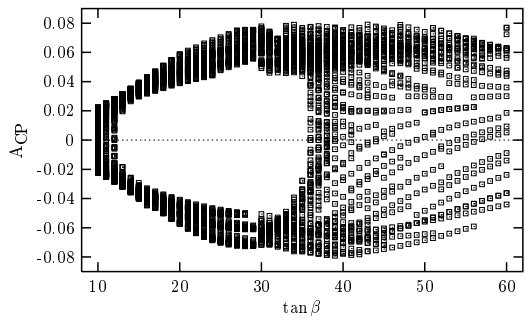}}
\centerline{\parbox{15cm}{\caption{\label{fig:Boz}
Direct CP violation in a SUSY model with minimal flavor violation but 
explicit CP violation \cite{Boz:2002wa}.}}}
\end{figure}

Large CP asymmetries are possible in many extensions of the SM with new 
CP-violating couplings entering the Wilson coefficients. A useful 
approximate expression for the asymmetry (assuming that there are no new 
operators present) is \cite{Kagan:1998bh}
\[
   A_{\rm CP}\approx 1.3\%\,\mbox{Im}(C_2/C_7)
   - 9.5\%\,\mbox{Im}(C_8/C_7) \,.
\]
The first term is important for models with $|C_7|\approx|C_7^{\rm SM}|$ 
but with a non-standard phase ($\mbox{arg}(C_7)\ne 0$), and can lead to 
CP asymmetries of about 5\%. The second term is important for models with 
enhanced chromo-magnetic dipole transitions ($C_8$) and new CP-violating 
couplings, and can lead to CP asymmetries exceeding 10--20\% without 
conflicting with the total $B\to X_s\gamma$ branching ratio. 
Figure~\ref{fig:Boz} illustrates this fact in the context of the MSSM with
minimal flavor violation but explicit CP violation 
($\phi_\mu,\phi_A\ne 0$) \cite{Demir:2001yz,Boz:2002wa}. Including 
large-$\tan\beta$ enhanced contributions beyond leading order, one finds 
significant complex contributions to $C_{7,8}$, which can lead to 
$A_{\rm CP}(B\to X_s\gamma)$ of order 10\% without spoiling the SM 
prediction for the branching ratio.

\section{Photon Spectrum as a QCD Tool}

The $B\to X_s\gamma$ photon-energy spectrum is insensitive to New-Physics 
effects and therefore a great QCD laboratory. It is useful for measuring
with good precision some hadronic parameters that are important elsewhere 
in $B$ physics and, in particular, for the determination of the CKM matrix. 
The moments $\langle E_\gamma\rangle$ and 
$(\langle E_\gamma^2\rangle-\langle E_\gamma\rangle^2)$ of the photon 
spectrum provide a precise determination of the $b$-quark mass and other 
heavy-quark effective theory parameters, which helps in the determination 
of $|V_{cb}|$. Combining information from the $B\to X_s\gamma$ photon 
spectrum and the $B\to X_u l\nu$ charged-lepton spectrum provides for the 
currently best route to measuring $|V_{ub}|$, which is immensely important 
for unitarity-triangle physics at the $B$ factories.

Let me illustrate this connection in a bit more detail. It has been shown 
long ago that the leading non-perturbative effects in the endpoint 
regions of the $B\to X_s\gamma$ photon spectrum and the 
$B\to X_u l\nu$ charged-lepton spectrum can be related to a universal 
shape function (up to $\Lambda_{\rm QCD}/m_b$ corrections) 
\cite{Neubert:1993um}. Using a measurement of the $B\to X_s\gamma$ photon 
spectrum $S(E_\gamma)$, one can predict the fraction of $B\to X_u l\nu$
events with charged-lepton energy $E_l>E_0$ via
\[
   F_u(E_0) = \int_{E_0}^{m_B/2}\!dE_\gamma\,w(E_\gamma,E_0)\,
   S(E_\gamma) \,,
\]
where the weight function $w(E_\gamma,E_0)$ is known including 
perturbative and $\Lambda_{\rm QCD}/m_b$ corrections
\cite{Neubert:1993um,Leibovich:1999xf,Bauer:2002yu,Neubert:2002yx}.
One can then extract $|V_{ub}|$ from a measurement of the 
$B\to X_u l\nu$ decay rate in the region above 2.2\,GeV. The resulting 
theoretical uncertainty on $|V_{ub}|$ is of order 10\% or less. The first 
experimental analysis using this strategy has been presented by CLEO 
\cite{Bornheim:2002du}. It gives the rather precise value 
\[
   |V_{ub}| = (4.08\pm 0.56_{\rm exp}\pm 0.29_{\rm th})\cdot 10^{-3} \,.
\]

\section{Exclusive Radiative Decays}

Folklore says that the exclusive decays $B\to K^*\gamma$ and 
$B\to\rho\gamma$ are affected by large hadronic uncertainties and so are 
not very useful as far as searches for New Physics are concerned. This 
is a misconception.

\subsection{QCD factorization}

There has been significant recent progress in the theory of exclusive 
hadronic $B$ decays based on QCD factorization theorems 
\cite{Beneke:1999br}. In particular, a factorization formula for 
$B\to V\gamma$ decays (with $V=K^*$ or $\rho$) has been established
\cite{Beneke:2001at,Bosch:2001gv}, which reads
\[
   \langle V\gamma(\epsilon)|Q_i|B\rangle = \left[ F_{B\to V}(0)\,T_i^I
   + \int_0^1\!d\xi\,dx\,T_i^{II}(\xi,x)\,\Phi_B(\xi)\,\Phi_V(x) \right]
   \cdot\epsilon^* \,.
\]
This formula is believed to be true to all orders in perturbation theory,
and up to corrections of order $\Lambda_{\rm QCD}/m_b$, which can be 
expected to be small. 

The establishment of QCD factorization as the leading term in a rigorous 
heavy-quark expansion opens up novel strategies for New-Physics searches, 
since e.g.\ the CP asymmetries in exclusive modes can be enhanced with 
respect to those in inclusive decays. A particularly important application 
of this formalism concerns the CKM-suppressed $b\to d\gamma$ transitions, 
where inclusive measurements are hindered by the large $b\to s\gamma$ 
background. The SM prediction is that $b\to d\gamma$ decays are about 20 
times smaller than the corresponding $b\to s\gamma$ decays,\footnote{This 
expectation is supported by the tight experimental bounds 
$\mbox{Br}(B^-\to\rho^-\gamma)<2.3\cdot 10^{-6}$ and 
$\mbox{Br}(B^0\to\rho^0\gamma)<1.4\cdot 10^{-6}$ reported by BaBar 
\cite{Aubert:2002pa}, which imply 
$\mbox{Br}(B^-\to\rho^-\gamma)/\mbox{Br}(B^-\to K^{*-}\gamma)<0.06$ and 
$2\mbox{Br}(B^0\to\rho^0\gamma)/\mbox{Br}(B^0\to K^{*0}\gamma)<0.07$.}
but CP asymmetries are predicted to be 20 times larger! 

\subsection{Photon polarization}

Radiative $B$ decays in the SM predominantly have helicity structure 
$b_R\to s_L\gamma_L$; however, in many extensions of the SM (left-right 
symmetric models, some SUSY models, etc.) there can be couplings with 
opposite helicity. It has been suggested that the photon polarization 
could be measured in exclusive decays of the type $B\to K_{\rm res}\gamma$ 
followed by $K_{\rm res}\to K^*\pi\to K\pi\pi$, by studying the up-down 
asymmetry of the photon direction relative to the $K\pi\pi$ decay plane 
\cite{Gronau:2001ng}. The resulting asymmetry has been calculated to be 
$(34\pm 5)\%$ for $K_1(1400)$. Gross deviations from this prediction could 
signal the presence of opposite-chirality transitions induced by physics
beyond the SM.

\subsection{Isospin violation in \boldmath$B\to K^*\gamma$ 
decays\unboldmath}

\begin{figure}
\epsfxsize=9.5cm
\centerline{\epsffile{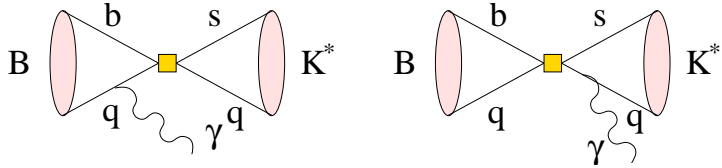}}
\centerline{\parbox{15cm}{\caption{\label{fig:iso}
Dominant SM source of isospin violation in $B\to K^*\gamma$ decays.}}}
\end{figure}

In the SM, the theoretical prediction for the isospin asymmetry
\cite{Kagan:2001zk}
\[
   \Delta_{0-} =
   \frac{\Gamma(B^0\to K^{*0}\gamma)-\Gamma(B^-\to K^{*-}\gamma)}
        {\Gamma(B^0\to K^{*0}\gamma)-\Gamma(B^-\to K^{*-}\gamma)}
   = (8\pm 3)\%
\]
is dominated by a contribution due to the penguin operator 
$Q_6=(\bar s_i b_j)_{V-A}\sum_q(\bar q_j q_i)_{V+A}$, as illustrated in
Figure~\ref{fig:iso}. As a result, this asymmetry is a direct probe of 
the sign and magnitude of the ratio Re$(C_6/C_7)$ of Wilson coefficients, 
thus providing a completely new window to New Physics (in the sense of 
probing a new operator). If future precise measurements could establish
a positive value for the asymmetry, as predicted by the SM, this would 
exclude a large portion of MSSM parameter space at large $\tan\beta$ 
\cite{Kagan:2001zk}.

\section{Conclusions}

Rare radiative decays based on the quark transition $b\to s\gamma$ are the
``mother'' of all FCNC processes. They benefit from good theoretical 
control and experimental accessibility, large sensitivity to New Physics, 
and the availability of many observables (rates, CP asymmetries, photon 
polarization, isospin violation). The present data already place tight 
constraints on several extensions of the SM (including SUSY models), but 
more detailed analyses exploring many observables are needed to thoroughly 
probe for New Physics.

In this talk I had no time to discuss other, related processes such as 
$b\to s l^+ l^-$, $b\to s\nu\bar\nu$, $K\to\pi\nu\bar\nu$, which are 
equally rich in their phenomenology and their reach for physics beyond the
SM. Although analyses of flavor-changing processes in radiative and other 
rare $B$ decays have so far not shown any evidence (within present errors) 
for physics beyond the SM, only a pessimist would use this fact as an 
argument against supersymmetry. An optimist would instead look forward to 
SUSY 2003!

\vspace{0.3cm}  
{\it Acknowledgment:\/} 
I am grateful to the organizers of SUSY 2002 for the invitation to present
this talk and for financial support. My attendance at this workshop was 
also supported by the National Science Foundation under Grant PHY-0098631.

\end{document}